
\magnification=1200
\hoffset=-.1in
\voffset=-.2in

\vsize=7.5in
\hsize=5.6in
\tolerance 10000

\baselineskip 13pt plus 1pt minus 1pt

\def\footnoterule{\kern-3pt \hrule width \hsize \kern6.2pt}
\def\pmb#1{\setbox0=\hbox{$#1$}%
\kern-.025em\copy0\kern-\wd0
\kern.05em\copy0\kern-\wd0
\kern-.025em\raise.0433em\box0 }

\def\lessim{\lower0.6ex\hbox{$\,$\vbox{\offinterlineskip
\hbox{$<$}\vskip1pt\hbox{$\sim$}}$\,$}}

\def\grtsim{\lower0.6ex\hbox{$\,$\vbox{\offinterlineskip
\hbox{$>$}\vskip1pt\hbox{$\sim$}}$\,$}}

\pageno=0

\footline={\ifnum\pageno>0 \hss --\folio-- \hss \else\fi}
\centerline{\bf STRING THEORY AND INFLATION}

\vskip 24pt

\centerline{Thibault Damour}
\vskip 12pt
\centerline{\it Institut des Hautes Etudes Scientifiques,}
\centerline{\it F-91440, Bures sur Yvette, France, and}
\centerline{\it DARC, CNRS-Observatoire de Paris,}
\centerline{\it F-92195 Meudon, France}
\vskip 12pt

\centerline{Alexander Vilenkin}
\vskip 12pt
\centerline{\it Institute of Cosmology}
\centerline{\it Department of Physics and Astronomy}
\centerline{\it Tufts University, Medford, MA 02155, USA}
\vskip 1.5in

\centerline{\bf ABSTRACT}
\medskip
String theory abounds with light scalar fields (the dilaton and various
moduli) which create a host of observational problems, and notably some serious
cosmological difficulties similar to the ones associated with the Polonyi field
in the earliest versions of spontaneously broken supergravity. We show that all
these problems are naturally avoided if a recently introduced mechanism [16]
for fixing the vacuum expectation values of the dilaton and/or moduli is at
work. We study both the classical evolution and the quantum fluctuations of
such scalar fields during a primordial inflationary era and find that the
results are naturally compatible with observational facts. In this model,
dilatons or moduli within a very wide range of masses (which includes the
SUSY-breaking favored $\sim 1$ TeV value and extends up to the Planck scale)
qualify to define a novel type of essentially stable ultra-weakly interacting
massive particles able to provide enough mass density to close the universe.

\vfill

\eject

\baselineskip 20pt plus 1pt minus 1pt

\noindent{\bf I.\quad INTRODUCTION}
\medskip
\nobreak

Superstring unification [1] and inflationary scenario [2] have been, arguably,
the most influential ideas in particle physics and cosmology over the last
decade.  Attempts at combining these ideas and constructing
superstring-based inflationary models have introduced some interesting
new ideas [3], but have, however, encountered serious difficulties
[4-7].
A specific source of difficulties is the existence of massless scalar
fields (the moduli) having only gravitational strength couplings to
ordinary matter. Among the moduli fields, the dilaton $\Phi$,
 which invariably accompanies the graviton in all superstring
models, plays a special role.
  In the presence of a (tree-level coupled) dilaton, the cosmological evolution
is drastically different from that in Einstein's gravity.  In particular,
instead of driving an exponential inflationary expansion, a constant vacuum
energy drives the dilaton towards large negative values
(corresponding to weak couplings), while the universe expands only as a
small power of time.
  Even apart from inflation, a massless dilaton, or moduli, field
 (for simplicity, as these fields share many characteristics,
we shall refer to any of them as `` the dilaton'' and denote it as $\Phi$)
 can cause a host of other cosmological problems.  During the most
recent, matter-dominated epoch of the universe, such field is necessarily
time-dependent.  The masses of elementary particles and their couplings all
depend on $\Phi$, and the cosmological time-variation of the dilaton runs into
a violent conflict with observations, not to mention unacceptably large
violations of the  equivalence principle.

These problems are usually addressed by assuming that the dilaton develops a
potential, so that one recovers the standard Einstein's gravity after $\Phi$
settles at the minimum of the potential.  The potential could originate from
non-perturbative effects, such as gaugino condensation.
The existence of a potential for $\Phi$ entails, however, new difficulties.
 On the one hand, it has been
argued [7] that the minimum of the nonperturbative potential is too shallow to
confine the dilaton without fine-tuning the initial conditions (see however
[8]).
On the other hand, any potential $V(\Phi)$ for a weakly coupled field
resurrects the
Polonyi problem [9-12]: either the energy stored in the coherent oscillations
of  the vacuum expectation value (VEV) of $\Phi$ does not dissipate before
now and exceeds the critical density needed to close the universe,
or the field decays before now , and thereby generically produces
an excessive amount of entropy. In the case of the nonperturbative
potentials suggested by current SUSY breaking models in string theory,
the slow decay rate of the moduli fields leaves the universe in a
radiation-dominated era at a temperature which is generically (i.e. under
naturalness assumptions for the couplings) much too low  to be consistent  with
primordial nucleosynthesis [13-15].

 In the present paper, we show that all the above difficulties
associated with the dilaton (and moduli) fields are naturally
avoided if the new mechanism introduced in Ref. [16] for fixing $\Phi$
is at work [17]. The main idea of [16] was to exploit the fact
 that string loop effects,
associated with worldsheets of arbitrary genus in intermediate string states,
naturally generate some non-monotonic dependence upon $\Phi$ of the
various couplings of $\Phi$ to the other fields. Under the assumption
that the different coupling functions $B_a(\Phi)$ have  extrema
at some common value of $\Phi = \Phi_0$, it has been
shown that interactions with massive particles
in an expanding universe drive the dilaton towards the value $\Phi_0$
at which it {\it decouples} from matter.
 All deviations from general relativity in this
model have been estimated to be  extremely small at the present cosmological
epoch thereby
 naturally reconciling a massless dilaton with existing observational data.
The assumption of coincident maxima can be satisfied  in a technically
natural way if there exists a discrete symmetry, e.g., $\Phi \to -\Phi$.   Such
a symmetry would guarantee that all couplings have extrema at $\Phi_0 = 0$
(the only additional necessary assumption being that these extrema lead to
minima rather than maxima of the masses as functions of $\Phi$).   We note
that precisely the discrete symmetry $\Phi \to -\Phi$  (or $e^\Phi \to
1/e^\Phi$) is known to hold for some of the moduli fields (T-duality), and has
been conjectured to hold for the  dilaton proper (S-duality: $g_s \to 1/g_s$,
where $g_s = e^\Phi$  is the string coupling).

Here we extend the analysis of Ref. [16] to inflationary models.
It will be shown that inflation is extremely efficient in driving a homogeneous
field $\Phi$ to $\Phi_0$ (Section II).  At the same time, inflation is known to
generate significant quantum fluctuations in gravitational and other fields
with a very wide range of wavelengths.  Fluctuations of the dilaton on
co-moving scales smaller than the present horizon are potentially dangerous
because they are not damped by the mechanism of Ref. [16].
One might worry that quantum fluctuations could reintroduce the Polonyi-moduli
problem in a different form. However, in Sec.III we shall
calculate the fluctuation spectrum and show that the predicted dilaton
fluctuations are well below the observational constraints.  We next consider,
in Sec.IV, the possibility that, in addition to non-trivial matter-coupling
functions with extrema at $\Phi = \Phi_0$, the dilaton also has a potential
with a minimum at $\Phi_0$.  (This may be enforced by the same discrete
symmetry).  In this case the dilatons are massive, and can potentially run
into conflict with observations, e.g. by overclosing the universe or by
generating an excessive flux of $\gamma$-rays due to dilaton decays.  We
shall see, however, that in our model the constraints on the dilaton mass
derived in [11-13], [18] can be substantially relaxed, due to the very weak
couplings of the dilaton.  We find that dilatons with a very large range of
masses  (which includes the suggested SUSY-breaking value \break $\sim$ TeV and
extends  up to the Planck scale)  can qualify to define a new type of
(essentially) stable WIMP able to provide enough mass density to
close the universe.

\bigskip

\noindent{\bf II.\quad THE EFFECTIVE ACTION}

\medskip

\nobreak

Up till now, most of the analysis of superstring cosmology has been based on
the tree-level effective action, corresponding to the lowest order in the
string loop expansion,
$$
S_{\rm tree} = \int d^4 x \sqrt {\hat g} e^{-2\Phi} \{ (\alpha')^{-1}
{\hat R} + 4{\hat
\nabla}^2 \Phi - 4({\hat \nabla}\Phi)^2 ] + {\cal L}_{\rm matter} \}.
\eqno{(2.1)}
$$
Here, $\Phi$ is the dilaton, and the matter Lagrangian includes fermions, gauge
and Higgs fields, and in particular the `inflaton' scalar field ${\hat \chi}$
whose potential ${\hat V}({\hat \chi})$ drives the inflation,
$$
{\cal L}_{\rm matter} = -{k \over{4}}{\hat F}^2 - {\bar {\hat \psi}}{\hat
D}{\hat \psi} - {1\over{2}}({\hat \nabla}{\hat \chi})^2 - {\hat V}({\hat
\chi}) +... . \eqno{(2.2)}$$
Hats in Eqs.(2.1), (2.2) indicate that the corresponding fields are taken in
the `string frame', that is, in the $\sigma$-model formulation of string
theory.

The string coupling $g_s$, which plays the role of the expansion parameter in
the string loop expansion, is determined by the expectation value of the
dilaton, $g_s = e^\Phi$.  The tree-level action (2.1) is proportional to
$g_s^{-2}$, resulting in a universal, multiplicative coupling of the dilaton to
all other fields.  With higher orders in the loop expansion taken into account,
we expect the common factor $e^{-2\Phi}$ to be replaced by several
coupling functions
$B_a (\Phi)$ multiplying different terms in (2.1), (2.2).  In particular, the
effective action for the graviton-dilaton-inflaton sector
is expected to be of the form
$$
S = \int d^4 x \sqrt{\hat g} \left\{ {B_g (\Phi) \over{\alpha'}}{\hat R} +
{B_\Phi (\Phi) \over{\alpha'}}[4{\hat \nabla}^2 \Phi - 4({\hat \nabla}\Phi)^2]
-
{1 \over{2}} B_\chi (\Phi) ({\hat \nabla}{\hat \chi})^2 - {\hat V}({\hat
\chi},\Phi) \right\} ,
\eqno{(2.3)}$$
where the functions $B_a (\Phi)$ admit a series expansion
$$
B_a (\Phi) = e^{-2\Phi} + c_0^{(a)} + c_1^{(a)} e^{2\Phi} + ... ,
\eqno{(2.4)}$$
and a similar expansion for ${\hat V}({\hat \chi}, \Phi)$.
In the case of the moduli fields (by contrast with the
four-dimensional dilaton proper)
the effective action has also the generic form (2.3), the only difference
being that the non-trivial $\Phi$-dependence is absent at tree level,
and arises at one loop and beyond.

A more convenient form of the action can be obtained  by a conformal
transformation from the `string-frame' metric ${\hat g}_{\mu\nu}$ to the
`Einstein-frame' metric
$$
g_{\mu\nu} = CB_g (\Phi){\hat g}_{\mu\nu} ,
\eqno{(2.5)}$$
and by replacing the dilaton field $\Phi$ by the variable
$$
\varphi = \int d\Phi \left[ {3 \over{4}}\left( {{B_g}' \over{B_g}} \right)^2 +
2{{B_\Phi}' \over{B_g}} + 2{B_\Phi \over{B_g}} \right] ^{1/2} .
\eqno{(2.6)}$$
This gives
$$
S = \int d^4 x \sqrt{g} \left[ {1 \over{4q}}R - {1 \over{2q}}(\nabla
\varphi)^2 - {1 \over{2}}F(\varphi)(\nabla \chi)^2 - V(\chi,\varphi)
\right] ,
\eqno{(2.7)}$$
where we have defined
$$
\chi = C^{-1/2} {\hat \chi} ,
\eqno{(2.8a)}$$
$$
F(\varphi) = B_\chi (\Phi)/B_g (\Phi) ,
\eqno{(2.8b)}$$
$$
V(\chi,\varphi) = C^{-2} B_g^{-2} (\Phi){\hat V} ({\hat \chi}, \Phi) .
\eqno{(2.8c)}$$
The constant $C$ in Eq.(2.5) is chosen so that the string units coincide with
Einstein units at the present cosmological epoch, $CB_g (\Phi_0) = 1$, and the
constant $q$ in (2.7) is related to the bare gravitational constant ${\bar G}$,
$q = 4\pi {\bar G} = C\alpha' /4$. As shown in [16], ${\bar G}$ is numerically
nearly identical to the observed Newtonian gravitational constant so that
$q = 4\pi / {m_p^2}$ with $m_p = 1.22 \times 10^{19}$ GeV.

 The minimal condition  required for the
mechanism of Ref. [16] to work is that all $B_a (\Phi)$ should have an extremum
at the same $\Phi = \Phi_0$.
When formulated within the context of inflationary models and
 in terms of the rescaled fields $\varphi$ and
$\chi$, this leads to  requiring that the potential
$V(\chi,\varphi)$ in Eq.(2.7) has
a minimum (as a function of $\varphi$) at $\varphi = \varphi_0$ for any fixed
value of $\chi$.  Here, $\varphi_0 = \varphi (\Phi_0)$.
A simple toy model which satisfies this condition is the case where higher
loops are supposed to respect the tree-level universality of the
dilaton couplings.  In this case all functions
$B_a (\Phi)$ in (2.3) are identical,
$$
B_a (\Phi) = B(\Phi) ,
\eqno{(2.9a)}$$
and
$$
{\hat V}({\hat \chi},\Phi) = B(\Phi){\hat V}({\hat \chi}).
\eqno{(2.9b)}$$
Then, in the Einstein-frame action (2.7), $F(\varphi) = 1$ and the
potential has a factorized form,
$$
S = \int d^4 x \sqrt{g} \left[ {1 \over{4q}}R - {1 \over{2q}}(\nabla
\varphi)^2 - {1 \over{2}}(\nabla \chi)^2 - B^{-1} (\varphi) V(\chi)\right] .
\eqno{(2.10)}$$

In this model $\varphi$ is attracted (both during inflation and the
subsequent matter-dominated era(s) ) toward the maxima of $B(\varphi)$.
Although the strong universality condition (2.9) may be too restrictive,
it is useful to have in mind the
action (2.10) which provides a very simple model  containing most of the
essential physics of the more general model (2.7).

\vfill\eject

\noindent{\bf III.\quad CLASSICAL EVOLUTION}

\medskip

\nobreak

We shall assume, as it is usually done in inflationary models, that the
potential $V(\chi,\varphi)$ in Eq.(2.7) has a minimum with $V(\chi_0 ,
\varphi_0) \approx 0$ and that it has a `slow-roll' region where it is a slowly
varying function of $\chi$,
$$
{{V_\chi}'}^2 \ll 12qV^2 ,
\eqno{(3.1a)}$$
$$
{(\sqrt{V})''}_{\chi\chi} \ll 3q\sqrt{V} ,
\eqno{(3.1b)}$$
where we recall that $q = 4\pi/m_p^2$.
At the same time, the coupling functions $B_a$ and the potential $V$ are not
expected to be slowly-varying functions of $\varphi$.  For example, in the
strong universality model (2.9) we expect $B(\varphi_0) \sim 1$ and
$$
\kappa \equiv - B''(\varphi_0)/B(\varphi_0) \sim 1
\eqno{(3.2)}$$
at the maximum of $B(\varphi)$. Note that, contrary to $\chi$ which has the
usual dimension of mass, $\varphi$ is a dimensionless variable whose expected
range of variation is of order unity.

For the initial conditions of the universe, we shall assume that the fields
$\varphi$ and $\chi$ are displaced from the minimum of $V(\chi, \varphi)$ at
$\{ \chi_0, \varphi_0 \}$ but are in its basin of attraction, at least in some
parts of the universe [19].  Then it is easy to see, qualitatively, how the
cosmological evolution will proceed.  The field $\varphi$, which corresponds to
the steep direction in $V(\chi, \varphi)$, will evolve on a much faster time
scale than $\chi$.  It will start oscillating about $\varphi = \varphi_0$,
these oscillations will be damped by the expansion of the universe, and the
universe will quickly settle into a quasi-exponential inflation driven by the
potential energy of the field $\chi$, ${\tilde V}(\chi) = V(\chi,\varphi_0)$.
The damping of $\varphi$-oscillations during inflation is very efficient, and
we expect that by the end of inflation the dilaton will be very close to
$\varphi_0$.

To describe this quantitatively, let us consider the
field equations for $\chi$ and $\varphi$
$$
\nabla(F(\varphi) \nabla \chi)-{\partial \over{\partial \chi}}V(\chi,\varphi)
 = 0,
\eqno{(3.3)}
$$
$$
\nabla^2 \varphi -{1\over 2} q {\partial F (\varphi) \over \partial
\varphi} (\nabla \, \chi)^2 -q {\partial \over{\partial
\varphi}}V(\chi,\varphi) = 0 . \eqno{(3.4)}
$$
During the slow-roll phase of inflation, the universe can be locally described
by a flat Robertson-Walker metric,
$$
ds^2 = - dt^2 + a^2(t)d{\bf x}^2 ,
\eqno{(3.5)}$$
with the expansion rate $H = {\dot a}/a$ given by
$$
H^2 ={1\over 3} \left[ 2q V(\chi ,\varphi) + \dot{\varphi}^2 + qF(\varphi)
\dot{\chi}^2 \right] \approx {2 \over{3}}q V(\chi, \varphi_0) . \eqno{(3.6)}
$$
  The spacetime is approximately de Sitter, with the
curvature
$$
R \approx 12H^2 \approx 8 q V (\chi ,\varphi_0) .
\eqno{(3.7)}
$$

To study the approach of $\varphi$ to $\varphi_0$, we expand $V(\chi,\varphi)$
in powers of $(\varphi - \varphi_0)$. For notational consistency with [16],
it is convenient to denote by $\beta_i$ the (positive) dimensionless
parameter measuring the curvature (with respect to $\varphi$) of the
 inflationary mass scale around the minimum $\varphi_0$:
$$
\Lambda_i(\chi,\varphi) \equiv  V^{1/4}(\chi,\varphi) \approx
    \Lambda_i(\chi,\varphi_0) [ 1+ {1 \over{2}}\beta_i(\varphi -
\varphi_0)^2]. \eqno{(3.8)}
$$
Here, the index $i$ stands for `inflation'.
 In the simple model (2.9), in which the
potential is factorized, $V(\chi,\varphi) = B^{-1} (\varphi) V(\chi)$,
one has $\beta_i = \kappa/4$, where $\kappa$ was introduced in equation (3.2)
above. In the general case, the value of $\beta_i$ depends on the physics
determining the mass scale $\Lambda_i$.
As discussed in Ref. [16] (see Eq. (4.6b) there),  if the hierarchy $\Lambda_i
\ll  m_p$ (which seems necessary in inflationary models) is due to
nonperturbative effects, one expects to have  $\beta_i \sim {\rm ln}
(\Lambda_{\rm string} / \Lambda_i)  \grtsim 1$ .

 Substituting this in
Eq.(3.4) and using (3.6), (3.7), we obtain
$$
(\nabla^2 - \xi_i R)\delta \varphi = 0 ,
\eqno{(3.9)}$$
where $\delta \varphi = \varphi - \varphi_0$ and
$$
\xi_i = {1\over 2} \beta_i .
\eqno{(3.10)}$$
We note that Eq.(3.9) has exactly the form of that for a massless,
non-minimally coupled field.  The curvature $R$ is a slowly-varying function of
time during inflation, and the effect of the curvature term in (3.9) is similar
to that of a positive mass squared term $m^2 \sim H^2$ for the
fluctuations of $\varphi$. Inflation is followed by a $\chi$-dominated
expansion when the inflaton field $\chi$ oscillates about the minimum of its
potential (say $V\approx {1\over 2} m_{\chi}^2 (\varphi) \chi^2 )$. During
this $\chi$-dominated period, $\delta \varphi$ will approximately satisfy,
after taking an average over the $\chi$ oscillations, an equation of the
same form as (3.9) but with a different value of $\xi_{\chi} \equiv {1\over
2} \beta_{\chi}$ (here $\beta_{\chi} \equiv [\partial^2 {\rm ln} \, m_{\chi}
(\varphi)/\partial \varphi^2 ]_{\varphi =\varphi_0}$ measures the
$\varphi$-curvature of the mass of the $\chi$ field; see the Appendix for a
more exact equation satisfied by $\delta \varphi$).

Neglecting the spatial gradients of $\varphi$
(which are rapidly suppressed by the cosmological expansion) we can finally
rewrite (3.9) in the form
$$
\delta {\ddot \varphi} + 3H\delta{\dot \varphi} + 12\xi_i H^2 \delta \varphi =
0 .
\eqno{(3.11)}$$

During the slow roll period, the field $\varphi (t)$ changes at a much faster
rate than $H(t)$, and we can solve (3.11) using a WKB-type {\it ansatz},
$$
\delta \varphi = e^{W(t)} ,
\eqno{(3.12)}$$
and assuming $|{\ddot W}| \ll {\dot W}^2$.  Substitution of (3.12) into (3.11)
gives a quadratic equation for ${\dot W}$ with the solution
$$
{\dot W}(t) = \left( -{3 \over{2}} \pm \sqrt{{9 \over{4}} - 12\xi} \right) H(t)
{}.
\eqno{(3.13)}
$$
Alternatively we can follow the method of Ref. [20] and write  the equation
describing the evolution
of $\varphi$ in terms of the parameter $p$ measuring
 the number of $e$-foldings:
$$
p = \ln {a(t) \over{a(t_i)}} = \int_{t_i}^t Hdt ,
\eqno{(3.14)}$$
where $t_i$ denotes the time at the onset of inflation. The latter equation
can be written in all eras of interest, inflation, $\chi$-dominated (in the
approximation where one averages over the $\chi$ oscillations),
radiation-dominated, and matter dominated.  When neglecting  $(\varphi '_p)^2$
with respect to $1$, this equation reads
$$
{2 \over {3}} \varphi''_{pp} +
(1-\lambda) \varphi'_p  =
 - (1- 3 \lambda) \beta (\varphi - \varphi_0) \eqno (3.15)
$$
where $\lambda $ is the ratio between pressure and energy density (i.e. $-1,
0$ or $1/3$  in vacuum-, $\chi$- or matter-, and radiation-dominated eras,
respectively), and where $\beta$ is the  parameter measuring the curvature of
the $\varphi$-dependent mass scale driving  the evolution of $\varphi$ in the
corresponding era (e.g. $\beta_i$  in the inflationary era  and $\beta_{\chi}$
in the $\chi$-oscillation  dominated era).

The equation (3.15)
is that of a damped harmonic oscillator.  The critical value of $\beta$
separating the overdamped-type  solution from the damped-oscillation-type  one
is $\beta_c = 3/8$ (in both the vacuum-dominated and the matter-dominated
cases). During inflation, the approach of $\varphi$ toward $\varphi_0$ is
oscillatory if  $\beta_i > \beta_c$, i.e. $\nu^2 < 0$, where we define
$$
\nu^2 \equiv 6( \beta_c - \beta_i) = {9 \over{4}} - 12\xi_i .
\eqno (3.16)
$$
{}From what we said above, we expect to be
in this regime ($\nu = i |\nu|$). Then
$$
\delta \varphi = A e^{-3p/2} \cos (|\nu |p + \delta ) ,
\eqno{(3.17)}
$$
where $A$ and $\delta$ are constants.
  The number of $e$-foldings during inflation is $p_i \grtsim 65$, so that if
initially $|\delta \varphi | \sim 1$, i.e. $A \sim 1$, then, by the end of
inflation,
$$
|\delta \varphi | \lessim e^{-3p_i/2}  \lessim 10^{-42} .
\eqno{(3.18)}$$

During $\chi$-domination the average pressure of
the oscillating $\chi$-field is nearly zero, the expansion law  is $a(t)
\propto t^{2/3}$, and the approach of $\varphi$ toward $\varphi_0$ is
described by Eq. (3.15) with $\lambda =0$ and $\beta = \beta_{\chi}$. This
yields, when  $\beta_{\chi} > \beta_c$, an additional factor of attraction of
$\delta \varphi$ toward zero of order $e^{-3p_{\chi}/4}$, where $p_{\chi}$ is
the number of  e-foldings during $\chi$-domination. [Note that in the simple
model (2.10) one has  $\beta_{\chi} = {1\over 2} \kappa =2\beta_i$.] The
energy of $\chi$-oscillations eventually thermalizes, and the universe enters
the radiation era with
 $a(t) \propto t^{1/2}$.
Since $\lambda = 1/3$ or $R = 0$  then,
 we see either from (3.15) or from (3.9) that $\delta \varphi$ essentially
stops  evolving during the radiation era independently of the value of
$\beta$.  During the
subsequent matter-dominated era, $\delta \varphi$ will be further attracted
toward $\varphi_0$ by an additional  factor $e^{-3p_m/4}$, where $p_m \sim 9$
is the number of e-foldings  separating us from the end of the radiation era.
Finally, the present value  of $\delta \varphi$ is expected to be $$
|\delta \varphi | \lessim  10^{-49} .
\eqno (3.19)$$
Note that the estimate (3.19) is independent of the precise values of $\beta_i$
and $\beta_{\chi}$ as long as they are both $> 3/8$.
A nonzero value of $\delta \varphi$ causes a number of potentially observable
deviations from general relativity [16]. However, all observable
non-Einsteinian effects are proportional to the square of $\delta \varphi$.
The present  observational bounds (from equivalence principle tests) are,
within the context of the model (2.9),
$$
\kappa |\delta \varphi |_{\rm obs} \lessim  5 \times 10^{-6}. \eqno (3.20)
$$
Even if we were to assume that $\beta_i$, or equivalently $\xi_i$,
is anomalously smaller than unity, it would  only be in the
extreme case where $\xi_i \lessim 10^{-2}$  that the attraction
factor due to inflation $ \sim e^{-4 \xi_i p_i}$ would not be
much smaller than unity.
We see that inflation is extremely efficient in driving a homogeneous,
classical
field $\varphi$ to $\varphi_0$. In the case , considered below, where the
field $\varphi$ has a potential  (sharing the discrete symmetry of
the coupling functions $B_a(\varphi)$ ), we conclude that we have here
a natural , non fine-tuned, solution to the Polonyi-moduli problem
as the VEV of $\varphi$ is left, at the end of inflation,
 very precisely at the place where it stores no potential energy.
However, as the change in the equation of state at the end of inflation can
result in copious creation of dilatons we must consider whether this can
regenerate a non trivial quasi-classical VEV for $\varphi$.
  This will be discussed in the next section.

\bigskip

\noindent {\bf IV.\quad QUANTUM CREATION OF DILATONS}

\medskip

\nobreak

Particle creation during and shortly after inflation can be studied using the
standard methods of quantum field theory in curved spacetime [21].  For a
massless scalar field with coupling to the curvature, as in Eq.(3.9), this has
been done by Ford [22] in the limit where the coupling to the curvature
is nearly conformal, $|\xi_i - 1/6| \ll 1$.  He assumed also
that the $\chi$-dominated period is very short, so that inflation is followed
by thermalization in about one Hubble time.  He found that the energy
density of created particles at the end of inflation ($t = t_*$) is
$$
\rho_\varphi (t_*) \sim 10^{-2} (6\xi_i -1)^2 H_*^4 ,
\eqno{(4.1)}$$
where $H_*$ is the expansion rate at $t=t_*$.

We have calculated the energy density and the spectrum of dilatons without
assuming $(\xi_i - 1/6 )$ to be small and without assuming that the
$\chi$-dominated period is necessarily short. We found that the matching to
a $\chi$-dominated expansion brings several qualitatively new features but
does not change drastically the overall quantitative results of a matching
to a radiation-dominated era. To keep things simple we discuss in the text
only the latter case.  The details of our calculation are given in the
Appendix (which contains also a brief discussion of the matching to a
$\chi$-dominated expansion), and here we shall only state the results.

The spectrum is expressed in terms of the co-moving wave number $k$, which is
equal to the physical momentum of the wave at $t = t_*$ (we set $a(t_*) =1$).
At later times the momentum is $p(t)=k/a(t)$, and the wavelength is $\lambda
(t) = (2\pi /k)a(t)$.  We find that
the energy spectrum of dilatons is peaked at $k \sim  H_*$ and
is exponentially suppressed for $k \gg H_*$.  In the limit of
long wavelengths, $k \ll H_*$,
$$
d\rho_\varphi (t) = {\Gamma^2 (\nu) \over{32\pi^3 a^4(t)}}
\left( \nu - {1 \over{2}} \right)^2
\left( {2H_* \over{k}} \right)^{2\nu +1} k^3 dk
\eqno{(4.2)}$$
for $\nu$ real (i.e. $\xi_i < 3/16$ in Eq. (3.16)) and
$$
d\rho_\varphi (t) = {H_* \over 8\pi^2 |\nu| a^4 (t) } \left( |\nu|^2 + {1
\over 4} \right) k^2 dk
\eqno{(4.3)}
$$
for $\nu = i|\nu|$, where in the latter case we assumed for simplicity that
$\exp (-\pi |\nu|) \ll 1$. [Note that $\xi_i = 1/6$ corresponds to
$ \nu = 1/2$ for which Eq. (4.2) indeed predicts no particle production.]

When inflation is followed by $\chi$-domination, the dilaton spectrum
contains a second peak at $k\sim m_{\chi}$ which contributes roughly the same
number density as the peak at $k\sim  H_*$ and an energy
density differing by a factor $\sim m_{\chi} /H_*$. By integrating the spectra
(4.2) or (4.3) up to $k_{\rm max} \sim  H_* $
 we find (roughly independently of $\nu$ as long as $|\nu - 1/2| \grtsim 1$)
that the total dilaton energy density is of order
$$
\rho_\varphi (t) \sim 10^{-2}  H_*^4 a^{-4}(t).
 \eqno{(4.4)}
$$
In other words, we find that when $6\xi_i - 1$ becomes $\grtsim 1$, the factor
$ (6\xi_i - 1)^2$ present in the nearly conformal case (4.1) saturates to
something of order unity.

Using the approximate conservation of the comoving entropy $\propto {\cal N}
T^3 a^3$ and Hubble's law at thermalization $(H_* \sim {\cal N}_*^{1/2} T_*^2
/m_p)$ we can write for the present energy density in massless dilatons
$$
\Omega_\varphi \equiv {\rho_\varphi \over{\rho_c}} \sim 10^{-2}\left({H_*
\over{m_p}}\right)^2 \left({{\cal N} \over{{\cal N}_*}} \right)^{1/3} \Omega_r
. \eqno{(4.5)}
$$
Here $\rho_c$ is the critical density, ${\cal N}$ is the number of spin
degrees of freedom in radiation and relativistic particles (at present), ${\cal
N}_*$ is its value at $t_*$ (i.e. at thermalization),
$\Omega_r = \rho_r /\rho_c = 4
\times 10^{-5} h^{-2}$, $\rho_r$ is the radiation density, $h \sim 1/2$ is the
Hubble  parameter, and $m_p$ is the Planck mass.
The characteristic wavelength of dilatons for the peak at $k\sim H_*$
 (when $\xi_i \grtsim 1/12$ [23]) is
$$
\lambda_c \sim 2 \pi H_*^{-1} Z_* \sim 4 (m_p /H_*)^{1/2} {\rm mm} ,
\eqno{(4.6)}
$$
where $Z_* \sim {\cal N}_*^{{1\over 12}} {\cal N}^{-{1\over 3}} (H_* m_p)^{1/2}
/T$, with $T=2.74 K = (0.83 {\rm mm})^{-1}$, is the redshift at $t=t_*$. The
presence of a $\chi$-oscillatory era would add a (possibly overlapping)
second peak with
characteristic wavelength differing by a factor $\sim H_* /m_{\chi}$.

Waves of wavelength smaller than the Hubble length,
$\lambda (t) < t$, can be treated
classically, and their amplitude can be estimated from
$$
\varphi_k (t) \sim a(t) \left[ {q \over{k}}{d\rho_\varphi (t) \over{dk}}
\right]^{1/2} ,
\eqno{(4.7)}$$
and at the present time we find, using Eqs.(4.2), (4.3)
$$
\varphi_k \sim 0.1 q^{1/2} Z_*^{-1} k (H_* /k)^{\nu +1/2} ,~~~~~~ \nu >0 ,
\eqno{(4.8)}$$
$$
\varphi_k \sim 0.1 Z_*^{-1} (qH_* k)^{1/2} , ~~~~~~ \nu = i|\nu| .
\eqno{(4.9)}$$
For $k \sim  H_*$, corresponding to wavelengths $\sim
\lambda_c$, both of these equations give
$$
\varphi_{H_*} \sim (T/m_p)(H_* /m_p)^{1/2} \sim 10^{-32} (H_* /m_p)^{1/2}
\lessim 10^{-34} , \eqno{(4.10)}
$$
where we have used the bound [2] $H_* \lessim 10^{-5} m_p$ on the rate of
inflation.  (Larger values of $H_*$ result in an excessive amount of relic
gravitational waves).  In the case of $\nu^2 < 1/4 ~~ (\xi_i >1/6)$
[including $\nu^2 < 0$, i.e. $\xi_i > 3/16$] the dilaton amplitude $\varphi_k
\propto \left\vert k^{{1\over 2}-\nu} \right\vert$ decreases towards longer
wavelengths. This means that for the wavelengths and time scales of relevance
to laboratory experiments the quantum-regenerated $\delta
\varphi$ is many orders of magnitude below the observational bounds (3.20).

A different behavior is obtained for $\nu > 1/2 ~~ (\xi_i <1/6)$, when
$\varphi_k$ grows towards longer wavelengths.  The largest growth occurs
for nearly minimal coupling, $\xi_i \approx
0$, $\nu \approx 3/2$, when $\varphi_k \propto k^{-1 +4\xi_i}$, and on the
present Hubble scale $k\sim Ha \sim HZ_*$ with $H^{-1} \sim 10^{28}$ cm
$$
\delta \varphi_{\max} \sim \Omega_r^{1/2} (H/T)^{4\xi_i} (H_* /m_p)^{1-2\xi_i}
\sim 10^{-7} 10^{-106\xi_i} (10^5 H_* / m_p)^{1-2\xi_i} . \eqno{(4.11)}
$$
Even for anomalously small curvature couplings $\xi_i \lessim 10^{-2}$, and
the maximal allowed value of $H_*$, the dilaton amplitude is smaller than
$\delta \varphi \sim 10^{-7}$, and therefore smaller than the observational
limits (3.20) which become less stringent when $\kappa = 8\xi_i$ is itself
small, i.e. when the interaction of dilatons with matter is suppressed due
to a small value of $\xi_i$.

It should be noted that Eqs.(4.7)-(4.9) cannot be used for wavelengths longer
than the Hubble length, where real particles cannot yet be distinguished from
vacuum polarization effects.  In particular, we cannot conclude from (4.8) that
$\varphi_k$ can become arbitrarily large in the limit of long wavelengths.  To
estimate the dispersion of the dilaton field on super-horizon scales, we
calculated the quantum expectation value $\langle\varphi^2 \rangle$.  The
calculation is outlined in the Appendix, and the result is that for $\nu >
1/2$ we have  $\langle\varphi^2 \rangle (t) \sim
\varphi_{k_H}^2$, where $k_H$ corresponds to the Hubble scale at time $t$.
This indicates that super-horizon wavelengths do not significantly contribute
to $\delta \varphi$.

\bigskip

\noindent{\bf V.\quad MASSIVE DILATONS}

\medskip

\nobreak

Up till now we have been considering the case where the dilaton remains
exactly massless at low energy. However, as we remarked above, under the
assumption of a discrete $\varphi$-symmetry (or some other universality
feature such as the one built in the model (2.9)) the existence of a mass
term for $\varphi$ does not create the usual Polonyi problem because, after
inflation, the VEV of $\varphi$ is left very precisely pinned at the minimum
of its potential. We must, however, investigate what constraints on the dilaton
mass $m_\varphi$ are obtained by requiring that the present mass density of
the quantum-generated dilatons does not exceed the critical density $\rho_c$.
To simplify the discussion, we shall consider only the case of $\nu^2 < 1/4$,
when long-wavelength modes with $k \ll H_*$ are unimportant, and the particle
interpretation of the field $\delta \varphi$ becomes valid shortly after
$t_*$.  The mass $m_\varphi$ and the dilaton number density $n_\varphi$ should
then satisfy the condition
$$
\Omega_{\varphi} \equiv {n_\varphi m_\varphi \over \rho_c} \lessim 1 .
\eqno{(5.1)}
$$
To estimate $n_\varphi$, we note that the ratio
$$
r =
n_\varphi /n_r , \eqno{(5.2)}
$$
where $n_r$ is the density of particles with
masses smaller than the temperature, remains approximately constant in the
course of cosmological evolution (assuming that the dilaton lifetime exceeds
the age of the universe, see below).  At the end of inflation, we find from
integrating the number density spectrum (see Appendix)
$$
n_\varphi (t_*) \sim 10^{-2}  H_*^3 .
\eqno{(5.3)}
$$
 Using the other equations
$$
n_r (t_*) \sim {\cal N}_* T_*^3 ,
\eqno{(5.4)}
$$
$$
H_* \sim {\cal N}_*^{1/2} T_*^2 /m_p ,
\eqno{(5.5)}
$$
where $T_*$ is the thermalization temperature, we get
$$
r \sim 10^{-2} {\cal N}_*^{-{1\over 4}} (H_* /m_p)^{3/2} .
\eqno{(5.6)}$$
Inserting this in (5.1) we obtain
$$
\Omega_{\varphi} \sim 10^{-2} {\cal N}_*^{-{1\over 4}} (H_* /m_p)^{3/2}
(m_{\varphi} /T) \Omega_r , \eqno (5.7a)
$$
where, as above, $\Omega_r = \rho_r /\rho_c = 4\times 10^{-5} h^{-2}$.
Numerically this reads
$$
\Omega_{\varphi} \sim \left( 10^5 {H_* \over m_p} \right)^{{3\over 2}}
{m_{\varphi} \over 10{\rm GeV}}. \eqno (5.7b)
$$

When the equality $\Omega_{\varphi} =1$ is satisfied, dilatons dominate the
mass density of the universe. This can happen for the whole range of masses
$m_{\varphi} \grtsim 10{\rm GeV}$. Let us note in particular that the value
$m_{\varphi} \sim 1{\rm TeV}$ suggested by many current SUSY breaking models
[13], [14] is allowed and corresponds to $H_* \sim 10^{-7} m_p$, i.e. $T_* \sim
10^{15} {\rm GeV}$. We remark also that the value $m_{\varphi} \sim m_p$ [24]
corresponds to a ``weak scale inflation'' $H_* \sim 100 {\rm GeV}$ [17] (i.e.
to an intermediate scale reheating temperature
$T_* \sim 3 \times 10^{10} {\rm GeV} \sim (m_W m_p)^{1/2}$).
Other authors have suggested the possibility that dilatons
may provide the dark matter of the universe [18], [25]. An important difference
of our model is that our allowed mass range is $m_{\varphi} \grtsim 10{\rm
GeV}$ (which contains notably $m_{\varphi} \sim 1 {\rm TeV}$) and
still corresponds to essentially {\it stable} dilatons, with a decay time much
larger than the age of the universe (as is discussed next). [Usually [18],
stable dilatons exist only for $m_{\varphi} \lessim 100 {\rm MeV}$ because of
Eq. (5.9) below.]

Let us finally examine whether additional constraints on the dilaton mass
follow from an eventual flux of $\gamma$-rays resulting from the dilaton decay,
$\varphi \to \gamma \gamma$.  This decay is described by the term $$
{\cal L}_{\rm int} \propto (\varphi - \varphi_0 ) F_{\mu \nu}^2
\eqno{(5.8)}
$$
in the effective Lagrangian. When expressed in terms of a canonically
normalized scalar field $\varphi_{\rm can} = \varphi / \sqrt q$, Eq. (5.8)
contains a coupling constant $\propto 1/m_p$. By dimensional analysis, if the
dimensionless coefficient in front of (5.8) is of order unity, the
corresponding lifetime is
$$
\tau_\varphi \sim m_p^2 / m_\varphi^3 .
\eqno{(5.9)}
$$

In our case, however, all coupling functions are expected to have extrema at
$\varphi = \varphi_0$, and thus the dimensionless coefficient in front of
(5.8) is $\propto \delta \varphi$, i.e. is exceedingly small. Then
$\gamma$-rays can only be produced in binary collisions $\varphi \varphi \to
\gamma \gamma$.  The corresponding interaction term is
$$
{\cal L}_{\rm int}
\propto (\varphi - \varphi_0 )^2 F_{\mu \nu}^2 , \eqno{(5.10)}
$$
with a dimensionless coefficient of order unity. The corresponding annihilation
cross-section is extremely small [26],
$$
\sigma \sim m_\varphi^2 /vm_p^4 ,
\eqno{(5.11)}
$$
where $v$ is the average velocity of dilatons.  The lifetime $\tau_\varphi$ can
be found from $n_\varphi \sigma v \tau_\varphi \sim 1$, which gives
$$
\tau_\varphi \sim m_p^4 /n_\varphi m_\varphi^2 .
\eqno{(5.12)}$$
The annihilation rate per unit spacetime volume is
$$
{n_\varphi \over{\tau_\varphi}} \sim {n_\varphi^2 m_\varphi^2 \over{m_p^4}}
\lessim {\rho_c^2 \over{m_p^4}} \sim t_0^{-4} ,
\eqno{(5.13)}$$
where $t_0$ is the present age of the universe.  Hence, there is no more than
one annihilation in the entire visible universe during its whole lifetime!

Note that as all coupling functions have extrema at $\varphi =\varphi_0$, the
type of dark matter our mechanism leads to has only exceedingly weak
interactions with ordinary matter. Basically, its presence can be felt only
through the gravitational effect of its mass. This leaves little hope of
detecting it in laboratory experiments. For instance, for a macroscopically
sizable mass $m_{\varphi} \sim m_p \sim 2\times 10^{-5} {\rm g}$, the average
present cosmological density of dilatons would be at most $n_{\varphi} \sim
10^{-24} {\rm cm}^{-3}$.

\bigskip
\noindent {\bf ACKNOWLEGEMENTS}
\medskip
\nobreak

T.D. thanks I. Antoniadis, M. Gasperini and G. Veneziano for informative
discussions. A.V. is grateful to Larry Ford and Keith Olive for discussions.
He also wishes to thank the Institut des Hautes Etudes Scientifiques for
hospitality during the period when most of this work was done
and the National Science Foundation for partial
support.

\bigskip
\noindent {\bf APPENDIX}
\medskip
\nobreak

In this Appendix we study the spectrum of created dilatons in  simple
models in which de Sitter inflation is followed either
by a radiation-dominated expansion or by a $\chi$-dominated one.
We consider first the transition to a radiation-dominated era.
 Using the conformal time coordinate $d\eta = dt/a(t)$, the
corresponding metric can be written as
$$
ds^2 = a^2(\eta )(-d\eta^2 + d{\bf x}^2 ) ,
\eqno{(A.1)}
$$
$$
a(\eta) = -(H_* \eta )^{-1} , ~~~~~~ \eta < \eta_* ,
\eqno{(A.2a)}$$
$$
a(\eta )= H_* (\eta - {\bar \eta}) , ~~~~~~ \eta > \eta_* .
\eqno{(A.2b)}$$
Here, $H_* = const$ is the expansion rate during inflation, $\eta_* <0$ is the
thermalization time when inflation ends, and ${\bar \eta} =\eta_* +(H_*^2
\eta_* )^{-1}$.  It will be convenient to set $\eta_* = -H_*^{-1}$, so that
$a(\eta_*)=1$.  The calculation in this Appendix follows the standard
techniques reviewed in [14].

The field operator $\delta {\hat \varphi}(x) = {\hat \varphi}(x)-\varphi_0$
satisfies a massless, non-minimally coupled field equation (3.9),
$$
(\nabla^2 -\xi_i R) \delta{\hat \varphi}(x) =0 ,
\eqno{(A.3)}$$
and can be expanded in terms of creation and annihilation operators,
$$
\delta {\hat \varphi}(x) = {q^{1/2} \over a(\eta)} \int {d^3k
\over{(2\pi)^{3/2}}} \left[ {\hat a}_{{\bf k}} \psi_k (\eta) e^{i{\bf kx}} +
h.c. \right] . \eqno{(A.4)}$$
Here, hats indicate operator quantities and should not be confused with the
notation of Section II where they indicate quantities in the `string frame'.
The mode functions $\psi_k (\eta)$ satisfy the normalization condition
$$
{\psi_k}' \psi_k^* - \psi_k {\psi_k^*}' = -i ,
\eqno{(A.5)}
$$
corresponding to $[\hat{a}_{{\bf k}} , \hat{a}_{{\bf k}'}^{\dagger}] = \delta
({\bf k} - {\bf k}')$. We shall assume that the quantum state of the dilaton
field during the inflationary period $\eta < \eta_*$ is the de Sitter-invariant
Bunch-Davis vacuum (i.e. that $\psi_k (\eta) \sim e^{-ik\eta} / \sqrt{2k}$
when $\eta \rightarrow -\infty$).  The corresponding mode functions are $$
\psi_k (\eta) = A(-\eta)^{1/2} H_\nu^{(1)} (-k\eta ) , \eqno{(A.6)}$$
where
$$
A= (\pi /4)^{1/2} e^{i\pi /4} e^{i\pi \nu /2} ,
\eqno{(A.7)}$$
$H_\nu^{(1)} (z)$ are Hankel functions, and $\nu$ is given by Eq.(3.16).

The mode functions for the radiation-dominated period $\eta > \eta_*$ are
$$
\psi_k (\eta)= {1 \over{(2k)^{1/2}}} \left[ \alpha_k
e^{-ik(\eta -{\bar \eta})} + \beta_k e^{ik(\eta -{\bar \eta})} \right] ,
\eqno{(A.8)}
$$
and Eq.(A.5) gives a normalization condition for $\alpha_k$ and $\beta_k$,
$$
|\alpha_k|^2 - |\beta_k|^2 = 1 .
\eqno{(A.9)}$$
The coefficients $\alpha_k$ and $\beta_k$ can be determined by matching the
mode functions (A.6) and (A.8) and their derivatives at $\eta = \eta_*$.  The
dilaton spectrum (in number density and energy density) can then be found from
$$
dn_\varphi ={1 \over{2\pi^2 a^3 (t)}} |\beta_k|^2 k^2 dk ,
\eqno{(A.10a)}
$$
$$
d\rho_\varphi ={1 \over{2\pi^2 a^4 (t)}} |\beta_k|^2 k^3 dk .
\eqno{(A.10b)}
$$

The instantaneous thermalization model (A.2) is not adequate for modes with
wavelengths shorter
than the de Sitter horizon at $\eta_*$, $k \grtsim H_*$.  In a more
realistic model, the transition from the vacuum to the radiation equation of
state takes at least a Hubble time, and particle creation in such modes is
exponentially suppressed, $\beta (k \gg H_* ) \approx 0$.  In the opposite
limit, $k \ll H_*$, we can find $\alpha_k$ and $\beta_k$ using the asymptotic
form of the Hankel functions for small values of the argument,
$$
H_\nu^{(1)} (z) \approx {i \over{\sin (\nu \pi )}} \left[ {e^{-i\nu \pi}
\over{\Gamma (1 +\nu)}} \left( {z \over{2}} \right)^\nu - {1 \over{\Gamma
(1-\nu )}} \left( {z \over{2}} \right)^{-\nu} \right] .
\eqno{(A.11)}$$
For $\nu >0$, the first term in (A.11) is negligible, and we find $\alpha_k
\approx -\beta_k$ and
$$
|\beta_k|^2 ={1 \over{16\pi}} \left( \nu -{1 \over{2}} \right)^2 \Gamma^2 (\nu
)
\left({2H_* \over{k}} \right)^{2\nu +1} .
\eqno{(A.12)}$$
In deriving (A.12) we have used the identity
$$
\Gamma (z) \Gamma (1-z) ={\pi \over{\sin ( \pi z )}} .
\eqno{(A.13)}$$

In the case of imaginary $\nu$, $\nu = i|\nu |$, we shall assume for simplicity
that $\exp (-\pi |\nu |) \ll 1$.  Then again the first term in (A.11) can be
neglected, and we obtain $\alpha_k \approx -\beta_k$ with
$$
|\beta_k |^2 \approx {H_* \over{4k |\nu |}} \left( |\nu |^2 +{1 \over{4}}
\right) .
\eqno{(A.14)}$$
Combining (A.12) and (A.14) with (A.10) we obtain Eqs.(4.2),(4.3). Let us
note in passing that the spectrum (A.14) is formally of the Rayleigh-Jeans
form: $\vert \beta_k \vert_{{\rm RJ}}^2 = T_{{\rm RJ}}/k$, that is the limit
of $(e^{k/T_{{\rm RJ}}} -1)^{-1}$ when $k\ll T_{{\rm RJ}}$. (Similar spectra
also come out of the superinflationary scenario of Refs. [3], [18]).
By contrast, we shall see below that when inflation is followed by
$\chi$-domination one does not get a Rayleigh-Jeans-type spectrum.

The dispersion of the dilaton field on super-horizon scales , $k \ll
\eta^{-1}$, can be estimated by calculating the expectation value
$$
\langle(\delta {\hat \varphi} )^2 \rangle = (2\pi )^{-3} a^{-2} q\int |\psi_k
(\eta )|^2 d^3 k . \eqno{(A.15)}
$$
The contribution of long wavelengths ($k \ll H_*$) to this integral can be
found using Eqs.(A.8),(A.12) and (A.14) with $\alpha_k \approx -\beta_k$. This
gives
$$
\langle (\delta {\hat \varphi} )^2 \rangle \approx {q \over{4\pi^3 H_*^2 (\eta
-{\bar \eta}  )^2}} \int d^3k k^{-1} |\beta_k |^2 \sin^2 [k(\eta -{\bar
\eta})] , \eqno{(A.16)}$$
and for $\nu >0$ we find
$$
\langle(\delta {\hat \varphi })^2 \rangle \sim q {(\nu -1/2)^2 \Gamma^2 (\nu )
\over{16\pi^3 H_*^2 (\eta -{\bar \eta})^2}}(2H_* )^{2\nu +1} \int_0^{\sim H_*}
dk k^{-2\nu} \sin^2 [k(\eta -{\bar \eta})] .
\eqno{(A.17)}$$
It is easily seen that for $\nu < 1/2$ the integral is dominated by the upper
limit ($k \sim H_*$), while for $1/2 < \nu < 3/2$ the dominant contribution is
given by $k \sim \eta^{-1}$, that is, by the modes of wavelength comparable to
the Hubble length.  In the latter case the upper limit of integration can be
extended to infinity, and we obtain a somewhat unwieldy expression
$$
\langle(\delta {\hat \varphi })^2 \rangle \sim -2\pi^{-3} \sin (\nu \pi )
\Gamma (1-2\nu ) \Gamma^2 (\nu) (\nu -1/2)^2 qH_*^2 [4a(\eta)]^{2\nu -3} ,
\eqno{(A.18)}$$
where $a(\eta)$ is given by (A.2b).  Disregarding numerical factors, this gives
$$
\langle(\delta {\hat \varphi})^2 \rangle \sim qH_*^2 [a(\eta)]^{2\nu -3} .
\eqno{(A.19)}$$

This is to be compared with the dilaton amplitude $\varphi_k$ which can be
found from Eqs.(4.7),(4.2),
$$
\varphi_k^2 (t) \sim {qk^2 \over{a^2 (t)}} \left({H_* \over{k}}
\right)^{2\nu +1} .
\eqno{(A.20)}$$
It is easily seen that $\langle(\delta {\hat \varphi})^2 \rangle \sim
\varphi_k^2$ for $k \sim \eta^{-1}$.

Let us finally briefly indicate the peculiarities of the more general case
where the inflationary evolution (A.2a) is followed by a $\chi$-dominated
expansion with an averaged scale factor (we assume $\Gamma_{\chi} \ll H(\eta)
 \ll m_{\chi}$ away from $\eta = \eta_*$
 and average over the oscillations at frequency $m_{\chi}$)
$$
\langle a(\eta) \rangle = \left[ 1+{1\over 2} H_* (\eta -\eta_* )\right]^2
\qquad (\eta > \eta_* ). \eqno (A.21)
$$
During this period, $\delta \varphi$ couples to the $\chi$-matter through the
$\varphi$-dependence of the potential energy $V(\chi ,\varphi) \approx
{1\over 2} m_{\chi}^2 (\varphi) \chi^2 \approx {1\over 2} m_{\chi}^2
(\varphi_0) \chi^2 \left[ 1+{1\over 2} \beta_{\chi} \delta \varphi^2 \right]^2$
which yields
$$
\nabla^2 \varphi - qV''_{\varphi \varphi} (\chi ,\varphi_0) \delta \varphi =
\nabla^2 \varphi - 2\beta_{\chi} q V(\chi ,\varphi_0) \delta \varphi = 0 .
\eqno (A.22)
$$
Then the Fourier modes $\psi_k (\eta)$ of Eq. (A.4) satisfy the
conformal-time evolution equation
$$
\partial_{\eta}^2 \psi_k (\eta) + (k^2 -U(\eta)) \psi_k (\eta) =0 , \eqno
(A.23)
$$
$$
U(\eta) \equiv {a''_{\eta \eta} \over a} -qa^2 V''_{\varphi \varphi} (\chi
,\varphi_0) = {a''_{\eta \eta} \over a} - 2q\beta_{\chi} a^2 V(\chi
,\varphi_0 ) . \eqno (A.24)
$$
The first forms of equations (A.22) and (A.24) are valid in both the
inflationary period (where $V''_{\varphi \varphi} (\chi ,\varphi_0) =
4\beta_i V(\chi ,\varphi_0)$) and the $\chi$-dominated one. During inflation
the effective potential reads $U(\eta) = 2(1-3\beta_i) \eta^{-2}$, while during
$\chi$-domination
$$
U(\eta) = {2\over (\eta -3\eta_*)^2} \left[1-3\beta_{\chi} +3(1-\beta_{\chi})
\cos [2m_{\chi} (t-t_*)]\right] \quad (\eta > \eta_*). \eqno (A.25)
$$
The effective potential (A.25) contains two distinct spectral features: a
monotonic piece $\propto (\eta -3\eta_*)^{-2}$ varying on the (averaged)
Hubble time scale, and an oscillatory  piece involving $\cos [2m_{\chi}
(t-t_*)]$. Correspondingly, the spectrum of quantum fluctuations generated by
solving the Schr\"odinger-like equation (A.23) will have two
peaks: one at $k\sim H_*$ and one at $k\sim m_{\chi}$.
[Depending upon the inflationary scenario considered, these peaks might be
separated or might overlap.] One can compute the
Bogolyubov coefficient $\beta_k$ of the first peak by matching at $\eta =
\eta_*$ the exact solutions of Eq. (A.23) when dropping the cosine term in
Eq. (A.25). As above, these are Hankel functions with $\nu =\nu_i \equiv i
\sqrt{6(\beta_i -3/8)}$ in the inflationary period and $\nu = \nu_{\chi}
\equiv i \sqrt{6(\beta_{\chi} -3/8)}$ during the $\chi$-dominated one. (We
assume here for definiteness that both are imaginary). This yields for the
``Hubble-time scale'' peak in the long wavelength limit $k\ll H_*$
$$
\vert \beta_k \vert_{{\rm H}}^2 = {1\over 2} {\vert \nu_{\chi}\vert \over
\vert \nu_i \vert} \left[ \left( {1\over 2} - {\vert \nu_i \vert \over \vert
\nu_{\chi}\vert}\right)^2 + \left( {3\over 4\vert \nu_{\chi}\vert} \right)^2
\right] . \eqno (A.26)
$$
Note that this $\vert \beta_k \vert_{{\rm H}}^2$ is independent of $k$, and
therefore different from the Rayleigh-Jeans-type spectrum (A.14). Integrating
(A.26) up to some effective cut-off $k_{\max} = \kappa H_* $ leads to
$$
n_{\varphi}^{{\rm H}} \sim {1 \over 48\pi^2} \kappa^3
{H_*^3 \over a^3} , \eqno (A.27a)
$$
$$
\rho_{\varphi}^{{\rm H}} \sim {1 \over 64\pi^2} \kappa^4
 {H_*^4 \over a^4} . \eqno (A.27b)
$$
The value of $\kappa$ depends upon the details of the transition between
inflation and $\chi$-domination. We expect that $\kappa \lessim 1$.
Assuming $\kappa \sim 1$, we get the total energy density (4.4) roughly
independently of whether inflation is followed by a radiation-dominated
era or a $\chi$-oscillatory one.

\noindent Finally, the ``oscillatory'' peak $k\sim m_{\chi}$ can be
estimated by applying Born perturbation theory to Eq. (A.23): namely,
$\beta_k^{{\rm osc}} \approx (2ik)^{-1} \int U^{{\rm osc}}
(\eta) e^{-2ik\eta} d\eta$. This yields ($\theta$ denoting the Heaviside step
function)
$$
\vert \beta_k \vert_{{\rm osc}}^2 = \left( {3\over 8}\right)^2 \pi
(\beta_{\chi} -1)^2 {H_*^3 m_{\chi}^{3/2} \over k^{9/2}} \theta (k-m_{\chi}),
\eqno (A.28)
$$
and, for instance,
$$
n_{\varphi}^{{\rm osc}} = {3\over 64\pi} (\beta_{\chi} -1)^2 {H_*^3 \over a^3}
. \eqno (A.29)
$$
[These estimates are accurate when $(\beta_{\chi} - 1) H_*^2 \ll m_{\chi}^2$
and should give the right order of magnitude when $\beta_{\chi} - 1 \sim 1$.]
In order of magnitude the number density (A.29) is comparable to (A.27a). The
corresponding energy density would differ from (A.27b) by a factor $\sim
m_{\chi} /H_*$.

\vfill
\eject

\centerline{\bf References}
\bigskip

\item{1}For a review of superstrings see, e.g., M.B. Green, J.H. Schwarz and
E. Witten, {\it Superstring Theory} (Cambridge U.P., Cambridge, 1987)
\medskip

\item{2}For a review of
inflation see A.D. Linde, {\it Particle Physics and Inflationary Cosmology}
(Harwood, Chur, 1990);  K.A. Olive, Phys. Rep. {\bf 190}, 307 (1990)
\medskip

\item{3}G. Veneziano, Phys. Lett. {\bf B265}, 287 (1991); M. Gasperini and
G. Veneziano, Astropart. Phys. {\bf 1}, 317 (1993)
\medskip

\item{4}P. Binetruy and M.K. Gaillard, Phys. Rev. {\bf D34}, 3069 (1986)
\medskip

\item{5}B.A. Campbell, A.D. Linde and K. Olive, Nucl. Phys. {\bf B355}, 146
(1991) \medskip

\item{6}A. Tseytlin and C. Vafa, Nucl. Phys. {\bf B372}, 443 (1992)
\medskip

\item{7}R. Brustein and P.J. Steinhardt, Phys. Lett. {\bf B302}, 196 (1993)
\medskip

\item{8}J.H. Horne and G. Moore, {\it Chaotic coupling constants},
hep-th/9403058
\medskip

\item{9}G.D. Coughlan, W. Fischler, E.W. Kolb, S. Raby and G.G. Ross, Phys.
Lett. {\bf B131}, 59 (1983)
\medskip

\item{10}A.S. Goncharov, A.D. Linde and M.I. Vysotsky, Phys. Lett. {\bf
B147}, 279 (1984) \medskip

\item{11}J. Ellis, D.V. Nanopoulos and M. Quiros, Phys. Lett. {\bf B174}, 176
(1986)
\medskip

\item{12}J. Ellis, N.C. Tsamis and M. Voloshin, Phys. Lett. {\bf B194}, 291
(1987)
\medskip

\item{13} B. de Carlos, J.A. Casas, F. Quevedo and E. Roulet, Phys. Lett. {\bf
B318}, 447 (1993)
\medskip

\item{14}T. Banks, D.B. Kaplan and A.E. Nelson, Phys. Rev. {\bf D49}, 779
(1994)
\medskip

\item{15}T. Banks, M. Berkooz and P.J. Steinhardt, Rutgers University Report,
RU-94-92

\item{16}T. Damour and A.M. Polyakov, Nucl. Phys. {\bf B423}, 532 (1994);
T. Damour and A.M. Polyakov, Gen. Rel. Grav. {\bf 26}, 1171 (1994)
\medskip

\item{17}After the completion of this work, our attention was drawn to a
different recent proposal for solving these difficulties. It consists of
invoking a brief period of ``weak-scale inflation''. See L. Randall and S.
Thomas, MIT-CTP-2331, NSF-IIP-94-70, SCIPP-94-16 (submitted to Nuc. Phys. B)
\medskip

\item{18}M. Gasperini and G. Veneziano, Phys. Rev. {\bf D50}, 2519 (1994); M.
Gasperini, Phys. Lett. {\bf B327}, 214 (1994)
\medskip

\item{19}Quantum cosmology suggests that inflating universes can spontaneously
nucleate out of nothing and provides a possible way of determining the initial
conditions in nucleating universes. [For a recent review and references see,
e.g., A.Vilenkin, Phys.Rev.{\bf 50}, 2581 (1994)].  For the model (2.7), the
nucleation is described by a de Sitter instanton with the fields $\varphi$ and
$\chi$ at one of the extrema of the potential $V(\varphi , \chi)$. The
instanton action, $|S| = 3m_p^4 /8V(\varphi ,\chi )$, is minimized by
{\it maximizing} $V(\varphi ,\chi )$.  For the simple model (2.10) this means
maximizing $V(\chi )$ and minimizing $B(\varphi )$.  Hence, according to
quantum cosmology, the universe is most likely to start at the highest
maximum of $V(\chi )$ and at the lowest minimum of $B(\varphi )$.  These are
just the right initial conditions for inflation.   \medskip

\item{20}T. Damour and K. Nordtvedt, Phys. Rev. Lett. {\bf 70}, 2217 (1993);
Phys. Rev. {\bf D48}, 3436 (1993)
\medskip

\item{21}N.D. Birrel and P.C.W. Davies, {\it Quantum Fields in Curved Space}
(Cambridge U.P., Cambridge, 1982).  For applications of these techniques to
inflation and references to the literature see Ref.[2].
\medskip

\item{22}L.H. Ford, Phys. Rev. {\bf D35}, 2955 (1987)
\medskip

\item{23}In the opposite case of nearly minimal coupling, $\xi_i \ll 1/12$, the
dilaton energy spectrum is nearly flat over a wide range of wavelengths (as in
the case of gravitons).
\medskip

\item{24}Our calculation of $n_\varphi$ in Sec.IV is valid only for
$m_\varphi^2 \ll 12 \xi_i H_*^2$.  However, in the opposite limit the
fractional change in $(m_\varphi^2 +\xi_i R )$ at the end of inflation is
small, and the production of $\varphi$-particles should be suppressed.  Hence,
we expect that the bound on $m_\varphi$ in this regime should be even weaker.
\medskip

\item{25}Instead of using dilatons as WIMPS, other authors have mentioned that
the Polonyi-type energy of oscillation of the VEV of $\varphi$ around the
minimum of its potential could close the universe. See e.g. Ref. [13] and a
recent preprint of P.J. Steinhardt and C.M. Will (Washington University
Report, WUGRAV-94-10).
\medskip

\item{26}Eq.(5.11) can be understood on dimensional grounds if we note that
each factor of $(\varphi - \varphi_0 )$ in the interaction Lagrangian gives a
factor of $m_p^{-1}$ in the scattering amplitude, and thus we should have
$\sigma \propto m_p^{-4}$.  The factor of $v^{-1}$ is a kinematic factor
appearing in all low-energy cross-sections.
\medskip

\par
\vfill
\end